\documentclass{aa}  
\usepackage{graphicx}
\usepackage{txfonts}

\usepackage{natbib}
\usepackage{color}
\usepackage{overpic}
\usepackage{pict2e} 
\begin{document} 

   \title{Excitation and charge transfer in low-energy hydrogen atom collisions with neutral iron \thanks{Data available in electronic form at the CDS via anonymous ftp to cdsarc.u-strasbg.fr (130.79.128.5) 
   or via http://cdsweb.u-strasbg.fr/cgi-bin/qcat?J/A+A/.  The data are also available at https://github.com/barklem/public-data.}}
   
   \titlerunning{Low-energy Fe+H collisions}


\author{P. S. Barklem\inst{1}}

   \institute{Theoretical Astrophysics, Department of Physics and Astronomy, Uppsala University,
              Box 516, SE-751 20 Uppsala, Sweden}
             
   \date{Received 27 Nov 2017 ; accepted 26 Dec 2017}

 
  \abstract
   {Data for inelastic processes due to hydrogen atom collisions with iron are needed for accurate modelling of the iron spectrum in late-type stars. Excitation and charge transfer in low-energy Fe+H collisions is studied theoretically using a previously presented method based on an asymptotic two-electron linear combination of atomic orbitals (LCAO) model of ionic-covalent interactions in the neutral atom-hydrogen-atom system, together with the multi-channel Landau-Zener model.  An extensive calculation including 166 covalent states and 25 ionic states is presented and rate coefficients are calculated for temperatures in the range 1000--20000~K.  The largest rates are found for charge transfer processes to and from two clusters of states around 6.3 and 6.6~eV excitation, corresponding in both cases to active $4d$ and $5p$ electrons undergoing transfer.  Excitation and de-excitation processes among these two sets of states are also significant.}


   \maketitle
%
\section{Introduction}
\label{sec:intro}

Spectral lines of neutral and singly ionised iron are so common in late-type stellar spectra that the iron abundance is used as a standard measure of overall metallicity and is often used as a diagnostic of stellar effective temperature and surface gravity via excitation and ionisation equilibria.  Thus, the iron spectrum can be used to derive the three most important parameters for late-type stellar atmospheres if its formation is well understood.  However, it has long been realised from both theoretical and observational evidence that the spectrum of neutral iron, \ion{Fe}{i}, departs from local thermodynamical equilibrium (LTE) \citep[e.g.][]{Athay1972, rutten_revision_1984, ruland_spectroscopic_1980, steenbock_statistical_1985, drake_fine_1991}.  The precise magnitude of the corrections due to departures from LTE has been a matter of significant debate, the efficiency of inelastic collision processes due to hydrogen leading to significant uncertainty in line formation calculations \citep[e.g.][]{holweger_solar_1996, thevenin_stellar_1999, gehren_kinetic_2001-1, Collet2005, mashonkina_non-lte_2011}.  Many studies have employed the Drawin formula, which is an extension of the classical Thomson model for ionisation by electrons to heavy particle collisions between like atoms \citep{Drawin1968}, including excitation \citep{Drawin1969,Drawin1973}, and is further extended to unlike atoms by \cite{steenbock_statistical_1984}.  This approach has been shown to be unsatisfactory for simple atoms, having poor agreement with experiment and full quantum calculations \citep[e.g.][]{Lambert1993,Barklem2011} and not capturing the dominant process, namely charge transfer.  However, for lack of better alternatives, this approach is still extensively used in the case of complex atoms such as iron, often with a scaling factor calibrated on spectra of standard stars.  This procedure is undesirable, as it is unclear that all discrepancies between theory and observation can be attributed to hydrogen collisions in the form described by the Drawin formula, and therefore doubtful that extrapolation of the calibration to other stars is accurate.

Detailed full quantum structure and scattering calculations of the type that have been employed for simple atoms \citep[e.g.][]{Belyaev1999,Belyaev2003,Barklem2003b,Belyaev2010,Barklem2010,Guitou2011,Belyaev2012,Barklem2012}, at the scale required for astrophysical applications, are currently beyond existing methods for complex atoms such as Fe.  In this paper, calculations for Fe+H inelastic collision processes are performed using a recently proposed asymptotic model approach \citep{barklem_excitation_2016, barklem_erratum:_2017}.  This approach employs a theoretical two-electron linear combination of atomic orbitals (LCAO) asymptotic model for the molecular structure, together with the multi-channel Landau-Zener model of nonadiabatic collision dynamics.  The model captures the ionic curve crossing mechanism seen to be important in experiment \citep{Fleck1991} and in the detailed calculations for simple atoms mentioned.  The method has been demonstrated to give reasonable estimates for the processes with large cross sections and rates when compared with detailed calculations, especially charge transfer processes.  Thus, these calculations aim to provide a more physically sound basis for understanding the role of collisions with hydrogen atoms in the formation of the iron spectrum in late-type stellar atmospheres.

\section{Calculations}

The calculations employ the method presented in \cite{barklem_excitation_2016, barklem_erratum:_2017} (hereafter B16) with updates as detailed in \cite{BarklemExcitationchargetransfer2017}.  These papers should be consulted for details of the model, including the notation for atomic parameters.  The aim was to include all states below the first ionic limit, i.e. below the asymptotic limit of the molecular state dissociating to $\mathrm{Fe}^+(a^6D) + \mathrm{H}^-$, which lies 57252~cm$^{-1}$ ($\sim$~7.1~eV, $\sim$~2.6~au) above the ground term of Fe+H.  A total of 166 states were included, up to and including $^5F^o$ at 56719~cm$^{-1}$, which has a crossing with the ionic state at over 400~$a_0$ and thus the crossing is practically diabatic.  The calculations are carried out in $LS$ coupling and states above this state might require an alternate coupling scheme, but in any case the crossings involving these states are practically diabatic.  

The energy level data employed were taken from National Institute of Standards and Technology (NIST) Atomic Spectra Database \citep{NIST_5.4} and are presented in Table~\ref{tab:input} (full table available electronically at CDS).  The required coefficients of fractional parentage are taken from standard tabulations \citep[e.g.][]{sobelman_atomic_1979}, and in the case of mixed configurations are combined using the method of \cite{Kelly1959}. In order to cover all ionic states appearing as core states among the included configurations, 25 ionic states were included as detailed in Table~\ref{tab:input}.  Many of these cores lead to crossings with covalent states only at very short internuclear distance, and could be excluded from the calculations without significant changes to the results; however, we choose to include these cores for completeness and this provides a check on the coefficients of fractional parentage (correct normalisation).  

The resulting molecular term symmetries are presented in Table~\ref{tab:syms}, including those 18 symmetries for which there are both ionic and covalent states, and thus are calculated.   The large number of ionic states (25 cores) and symmetries (18) lead to a large number of individual potential and dynamical calculations, although it should be noted that each ionic state has a limited number of symmetries.  As for previous calculations, cross sections are computed from thresholds to 100~eV collision energies, and rate coefficients are then calculated and finally summed over all symmetries and cores.  Final results are obtained for temperatures in the range 1000--20000~K with steps of 1000~K.

\begin{table*}
\center
\caption{\label{tab:input}  Input data for the calculations. Not all states are shown and the full table is available electronically at CDS.   The notation from the LCAO model in B16 is used, and detailed descriptions are given in that paper.  In short, $L_A$ and $S_A$ are the electronic orbital angular momentum and spin quantum numbers for the state of the iron atom, and $n$ and $l$ are the principal and angular momentum quantum numbers for the active electron. The value $E_j^\mathrm{Fe/Fe^+}$ is the state energy for the iron atom, $E_\mathrm{lim}$ the corresponding series limit, and $E_j$ the total asymptotic molecular energy.  The zero point in the case of energies on the iron atom, $E_j^\mathrm{Fe/Fe^+}$ and $E_\mathrm{lim}$, is the \ion{Fe}{i} ground term, and the zero point for the asymptotic molecular energies $E_j$ is the energy corresponding to both atoms in their ground states. The value  $N_\mathrm{eq}$ is the number of equivalent active electrons on the iron atom.The values $L_C$ and $S_C$ are the electronic orbital angular momentum and spin quantum numbers for the core of the iron atom;  $G^{S_A L_A}_{S_c L_c}$ is the coefficient of fractional parentage.   For covalent configurations in which iron is neutral and hydrogen is in the ground state, H($1s$) is implied and omitted for clarity. }
\begin{tabular}{lccccrrrclccr}
\hline \hline
$         \mathrm{Term}$ & $         L_A$ & $      2S_A+1$ & $           n$ & $           l$ & $           E_j^\mathrm{Fe/Fe^+}$ & $     E_\mathrm{lim}$ & $     E_\mathrm{j}$ & $      N_\mathrm{eq}$ & $          \mathrm{Core}$ & $         L_c$ & $      2S_c+1$ & $    G^{S_A L_A}_{S_c L_c}$ \\ 
  &  &  &  &  & [cm$^{-1}$] & [cm$^{-1}$] & [cm$^{-1}$] &  &  &  &  &  \\ \hline
  &  &  &  &  &  &  &  &  &  &  &  &  \\
\multicolumn{12}{c}{\underline{Covalent states}}  \\
$                                    a^5D$ & $           2$ & $           5$ & $           4$ & $           0$ & $           0$ & $       63335$ & $           0$ & $           2$ & $                       \mathrm{Fe}^+(a^6D)$  &$           2$ & $           6$ & $                    0.775$ \\
$                                    a^5D$ & $           2$ & $           5$ & $           4$ & $           0$ & $           0$ & $       71239$ & $           0$ & $           2$ & $                       \mathrm{Fe}^+(a^4D)$  &$           2$ & $           4$ & $                   -0.632$ \\
$                                    a^5F$ & $           3$ & $           5$ & $           4$ & $           0$ & $        7057$ & $       65335$ & $        7057$ & $           1$ & $                       \mathrm{Fe}^+(a^4F)$  &$           3$ & $           4$ & $                    1.000$ \\
$                                    a^3F$ & $           3$ & $           3$ & $           4$ & $           0$ & $       12004$ & $       65335$ & $       12004$ & $           1$ & $                       \mathrm{Fe}^+(a^4F)$  &$           3$ & $           4$ & $                    1.000$ \\
$                                    a^5P$ & $           1$ & $           5$ & $           4$ & $           0$ & $       17282$ & $       76531$ & $       17282$ & $           1$ & $                       \mathrm{Fe}^+(a^4P)$  &$           1$ & $           4$ & $                    1.000$ \\
$                                   a^3P2$ & $           1$ & $           3$ & $           4$ & $           0$ & $       18378$ & $       89088$ & $       18378$ & $           2$ & $                       \mathrm{Fe}^+(b^2P)$  &$           1$ & $           2$ & $                   -0.577$ \\
$                                   a^3P2$ & $           1$ & $           3$ & $           4$ & $           0$ & $       18551$ & $       84340$ & $       18551$ & $           2$ & $                       \mathrm{Fe}^+(b^4P)$  &$           1$ & $           4$ & $                    0.816$ \\
$                                    a^3H$ & $           5$ & $           3$ & $           4$ & $           0$ & $       19173$ & $       84952$ & $       19173$ & $           2$ & $                       \mathrm{Fe}^+(a^4H)$  &$           5$ & $           4$ & $                    0.816$ \\
$                                    a^3H$ & $           5$ & $           3$ & $           4$ & $           0$ & $       19173$ & $       89172$ & $       19173$ & $           2$ & $                       \mathrm{Fe}^+(b^2H)$  &$           5$ & $           2$ & $                   -0.577$ \\
$                                  z^7D^o$ & $           2$ & $           7$ & $           4$ & $           1$ & $       19221$ & $       63335$ & $       19221$ & $           1$ & $                       \mathrm{Fe}^+(a^6D)$  &$           2$ & $           6$ & $                    1.000$ \\
$                                   b^3F2$ & $           3$ & $           3$ & $           4$ & $           0$ & $       20411$ & $       86337$ & $       20411$ & $           2$ & $                       \mathrm{Fe}^+(b^4F)$  &$           3$ & $           4$ & $                    0.816$ \\
$                                   b^3F2$ & $           3$ & $           3$ & $           4$ & $           0$ & $       20411$ & $       90365$ & $       20411$ & $           2$ & $                       \mathrm{Fe}^+(a^2F)$  &$           3$ & $           2$ & $                   -0.577$ \\
$                                    a^3G$ & $           4$ & $           3$ & $           4$ & $           0$ & $       21546$ & $       78997$ & $       21546$ & $           1$ & $                       \mathrm{Fe}^+(a^2G)$  &$           4$ & $           2$ & $                    1.000$ \\
 ... & ... & ... & ... & ... & ... & ... & ... & ... & ... & ... & ... & ... \\
$                                     ^7F$ & $           3$ & $           7$ & $           5$ & $           2$ & $       56492$ & $       63335$ & $       56492$ & $           1$ & $                       \mathrm{Fe}^+(a^6D)$  &$           2$ & $           6$ & $                    1.000$ \\
$                                     ^5F$ & $           3$ & $           5$ & $           5$ & $           2$ & $       56500$ & $       63335$ & $       56500$ & $           1$ & $                       \mathrm{Fe}^+(a^6D)$  &$           2$ & $           6$ & $                    1.000$ \\
$                                     ^7G$ & $           4$ & $           7$ & $           5$ & $           2$ & $       56596$ & $       63335$ & $       56596$ & $           1$ & $                       \mathrm{Fe}^+(a^6D)$  &$           2$ & $           6$ & $                    1.000$ \\
$                                   ^5D^o$ & $           2$ & $           5$ & $           6$ & $           1$ & $       56704$ & $       63335$ & $       56704$ & $           1$ & $                       \mathrm{Fe}^+(a^6D)$  &$           2$ & $           6$ & $                    1.000$ \\
$                                   ^5F^o$ & $           3$ & $           5$ & $           6$ & $           1$ & $       56719$ & $       63335$ & $       56719$ & $           1$ & $                       \mathrm{Fe}^+(a^6D)$  &$           2$ & $           6$ & $                    1.000$ \\  
  &  &  &  &  &  &  &  &  &  &  &  &  \\
  \multicolumn{13}{c}{\underline{Ionic states}}  \\
$                               \mathrm{Fe}^+(a^6D) + \mathrm{H}^-$ & $           2$ & $           6$ & $           -$ & $           -$ & $       63335$ & $           -$ & $       57252$ & $           -$ & $                     $  &$           $ & $           $ & $                    $ \\
$                               \mathrm{Fe}^+(a^4F) + \mathrm{H}^-$ & $           3$ & $           4$ & $           -$ & $           -$ & $       65335$ & $           -$ & $       59252$ & $           -$ & $                     $  &$           $ & $           $ & $                    $ \\
$                               \mathrm{Fe}^+(a^4D) + \mathrm{H}^-$ & $           2$ & $           4$ & $           -$ & $           -$ & $       71239$ & $           -$ & $       65156$ & $           -$ & $                     $  &$           $ & $           $ & $                    $ \\
$                               \mathrm{Fe}^+(a^4P) + \mathrm{H}^-$ & $           1$ & $           4$ & $           -$ & $           -$ & $       76531$ & $           -$ & $       70448$ & $           -$ & $                     $  &$           $ & $           $ & $                    $ \\
$                               \mathrm{Fe}^+(a^2G) + \mathrm{H}^-$ & $           4$ & $           2$ & $           -$ & $           -$ & $       78997$ & $           -$ & $       72914$ & $           -$ & $                     $  &$           $ & $           $ & $                    $ \\
$                               \mathrm{Fe}^+(a^2P) + \mathrm{H}^-$ & $           1$ & $           2$ & $           -$ & $           -$ & $       81455$ & $           -$ & $       75372$ & $           -$ & $                     $  &$           $ & $           $ & $                    $ \\
$                               \mathrm{Fe}^+(a^2H) + \mathrm{H}^-$ & $           5$ & $           2$ & $           -$ & $           -$ & $       83471$ & $           -$ & $       77388$ & $           -$ & $                     $  &$           $ & $           $ & $                    $ \\
$                              \mathrm{Fe}^+(a^2D2) + \mathrm{H}^-$ & $           2$ & $           2$ & $           -$ & $           -$ & $       83752$ & $           -$ & $       77669$ & $           -$ & $                     $  &$           $ & $           $ & $                    $ \\
$                               \mathrm{Fe}^+(b^4P) + \mathrm{H}^-$ & $           1$ & $           4$ & $           -$ & $           -$ & $       84340$ & $           -$ & $       78257$ & $           -$ & $                     $  &$           $ & $           $ & $                    $ \\
$                               \mathrm{Fe}^+(a^4H) + \mathrm{H}^-$ & $           5$ & $           4$ & $           -$ & $           -$ & $       84378$ & $           -$ & $       78295$ & $           -$ & $                     $  &$           $ & $           $ & $                    $ \\
$                               \mathrm{Fe}^+(b^4F) + \mathrm{H}^-$ & $           3$ & $           4$ & $           -$ & $           -$ & $       85726$ & $           -$ & $       79643$ & $           -$ & $                     $  &$           $ & $           $ & $                    $ \\
$                               \mathrm{Fe}^+(a^6S) + \mathrm{H}^-$ & $           0$ & $           6$ & $           -$ & $           -$ & $       86236$ & $           -$ & $       80153$ & $           -$ & $                     $  &$           $ & $           $ & $                    $ \\
$                               \mathrm{Fe}^+(a^4G) + \mathrm{H}^-$ & $           4$ & $           4$ & $           -$ & $           -$ & $       88669$ & $           -$ & $       82586$ & $           -$ & $                     $  &$           $ & $           $ & $                    $ \\
$                               \mathrm{Fe}^+(b^2P) + \mathrm{H}^-$ & $           1$ & $           2$ & $           -$ & $           -$ & $       89088$ & $           -$ & $       83005$ & $           -$ & $                     $  &$           $ & $           $ & $                    $ \\
$                               \mathrm{Fe}^+(b^2H) + \mathrm{H}^-$ & $           5$ & $           2$ & $           -$ & $           -$ & $       89172$ & $           -$ & $       83089$ & $           -$ & $                     $  &$           $ & $           $ & $                    $ \\
$                               \mathrm{Fe}^+(a^2F) + \mathrm{H}^-$ & $           3$ & $           2$ & $           -$ & $           -$ & $       90365$ & $           -$ & $       84282$ & $           -$ & $                     $  &$           $ & $           $ & $                    $ \\
$                               \mathrm{Fe}^+(b^2G) + \mathrm{H}^-$ & $           4$ & $           2$ & $           -$ & $           -$ & $       93474$ & $           -$ & $       87391$ & $           -$ & $                     $  &$           $ & $           $ & $                    $ \\
$                               \mathrm{Fe}^+(b^4D) + \mathrm{H}^-$ & $           2$ & $           4$ & $           -$ & $           -$ & $       94338$ & $           -$ & $       88255$ & $           -$ & $                     $  &$           $ & $           $ & $                    $ \\
$                               \mathrm{Fe}^+(b^2F) + \mathrm{H}^-$ & $           3$ & $           2$ & $           -$ & $           -$ & $       94838$ & $           -$ & $       88755$ & $           -$ & $                     $  &$           $ & $           $ & $                    $ \\
$                               \mathrm{Fe}^+(a^2I) + \mathrm{H}^-$ & $           6$ & $           2$ & $           -$ & $           -$ & $       95810$ & $           -$ & $       89727$ & $           -$ & $                     $  &$           $ & $           $ & $                    $ \\
$                               \mathrm{Fe}^+(c^2G) + \mathrm{H}^-$ & $           4$ & $           2$ & $           -$ & $           -$ & $       96401$ & $           -$ & $       90318$ & $           -$ & $                     $  &$           $ & $           $ & $                    $ \\
$                               \mathrm{Fe}^+(b^2D) + \mathrm{H}^-$ & $           2$ & $           2$ & $           -$ & $           -$ & $       99121$ & $           -$ & $       93038$ & $           -$ & $                     $  &$           $ & $           $ & $                    $ \\
$                               \mathrm{Fe}^+(a^2S) + \mathrm{H}^-$ & $           0$ & $           2$ & $           -$ & $           -$ & $      100145$ & $           -$ & $       94062$ & $           -$ & $                     $  &$           $ & $           $ & $                    $ \\
$                               \mathrm{Fe}^+(c^2D) + \mathrm{H}^-$ & $           2$ & $           2$ & $           -$ & $           -$ & $      101103$ & $           -$ & $       95020$ & $           -$ & $                     $  &$           $ & $           $ & $                    $ \\
$                              \mathrm{Fe}^+(d^2D2) + \mathrm{H}^-$ & $           2$ & $           2$ & $           -$ & $           -$ & $      110812$ & $           -$ & $      104729$ & $           -$ & $                     $  &$           $ & $           $ & $                    $ \\

\hline
\end{tabular}
\end{table*}

\begin{table*}
\center
\caption{\label{tab:syms} Possible symmetries for Fe+H molecular states arising from various asymptotic atomic states and the total statistical weights.  The symmetries leading to covalent-ionic interactions among the considered states, and which thus need to be calculated, are shown at the bottom along with their statistical weights. For covalent configurations in which iron is neutral and/or hydrogen is in the ground state, this information is implied and omitted for clarity.   The full table is available electronically at CDS.}
\begin{tabular}{rlcl}
\hline \hline
Label & Term & $g_\mathrm{total}$& Molecular terms\\ \hline
$  1$ & $                                    a^5D$ &   50& $   ^{4}\Sigma^+,\        ^{4}\Pi,\     ^{4}\Delta,\   ^{6}\Sigma^+,\        ^{6}\Pi,\     ^{6}\Delta$ \\
$  2$ & $                                    a^5F$ &   70& $   ^{4}\Sigma^-,\        ^{4}\Pi,\     ^{4}\Delta,\       ^{4}\Phi,\   ^{6}\Sigma^-,\        ^{6}\Pi,\     ^{6}\Delta,\       ^{6}\Phi$ \\
$  3$ & $                                    a^3F$ &   42& $   ^{2}\Sigma^-,\        ^{2}\Pi,\     ^{2}\Delta,\       ^{2}\Phi,\   ^{4}\Sigma^-,\        ^{4}\Pi,\     ^{4}\Delta,\       ^{4}\Phi$ \\
$  4$ & $                                    a^5P$ &   30& $   ^{4}\Sigma^-,\        ^{4}\Pi,\   ^{6}\Sigma^-,\        ^{6}\Pi$ \\
$  5$ & $                                   a^3P2$ &   18& $   ^{2}\Sigma^-,\        ^{2}\Pi,\   ^{4}\Sigma^-,\        ^{4}\Pi$ \\
$  6$ & $                                   a^3P2$ &   18& $   ^{2}\Sigma^-,\        ^{2}\Pi,\   ^{4}\Sigma^-,\        ^{4}\Pi$ \\
$  7$ & $                                    a^3H$ &   66& $   ^{2}\Sigma^-,\        ^{2}\Pi,\     ^{2}\Delta,\       ^{2}\Phi,\     ^{2}\Gamma,\       ^{2}\mathrm{H},\   ^{4}\Sigma^-,\        ^{4}\Pi,\     ^{4}\Delta,\       ^{4}\Phi,\     ^{4}\Gamma,\       ^{4}\mathrm{H}$ \\
$  8$ & $                                  z^7D^o$ &   70& $   ^{6}\Sigma^-,\        ^{6}\Pi,\     ^{6}\Delta,\   ^{8}\Sigma^-,\        ^{8}\Pi,\     ^{8}\Delta$ \\
$  9$ & $                                   b^3F2$ &   42& $   ^{2}\Sigma^-,\        ^{2}\Pi,\     ^{2}\Delta,\       ^{2}\Phi,\   ^{4}\Sigma^-,\        ^{4}\Pi,\     ^{4}\Delta,\       ^{4}\Phi$ \\
$ 10$ & $                                    a^3G$ &   54& $   ^{2}\Sigma^+,\        ^{2}\Pi,\     ^{2}\Delta,\       ^{2}\Phi,\     ^{2}\Gamma,\   ^{4}\Sigma^+,\        ^{4}\Pi,\     ^{4}\Delta,\       ^{4}\Phi,\     ^{4}\Gamma$ \\
... & ...& ... & ... \\
$164$ & $                                     ^7G$ &  126& $   ^{6}\Sigma^+,\        ^{6}\Pi,\     ^{6}\Delta,\       ^{6}\Phi,\     ^{6}\Gamma,\   ^{8}\Sigma^+,\        ^{8}\Pi,\     ^{8}\Delta,\       ^{8}\Phi,\     ^{8}\Gamma$ \\
$165$ & $                                   ^5D^o$ &   50& $   ^{4}\Sigma^-,\        ^{4}\Pi,\     ^{4}\Delta,\   ^{6}\Sigma^-,\        ^{6}\Pi,\     ^{6}\Delta$ \\
$166$ & $                                   ^5F^o$ &   70& $   ^{4}\Sigma^+,\        ^{4}\Pi,\     ^{4}\Delta,\       ^{4}\Phi,\   ^{6}\Sigma^+,\        ^{6}\Pi,\     ^{6}\Delta,\       ^{6}\Phi$ \\
&&&\\
$167$ & $                      \mathrm{Fe}^+(a^6D)+\mathrm{H}^-$ &   30& $   ^{6}\Sigma^+,\        ^{6}\Pi,\     ^{6}\Delta$ \\
$168$ & $                      \mathrm{Fe}^+(a^4F)+\mathrm{H}^-$ &   28& $   ^{4}\Sigma^-,\        ^{4}\Pi,\     ^{4}\Delta,\       ^{4}\Phi$ \\
$169$ & $                      \mathrm{Fe}^+(a^4D)+\mathrm{H}^-$ &   20& $   ^{4}\Sigma^+,\        ^{4}\Pi,\     ^{4}\Delta$ \\
$170$ & $                      \mathrm{Fe}^+(a^4P)+\mathrm{H}^-$ &   12& $   ^{4}\Sigma^-,\        ^{4}\Pi$ \\
... & ...& ... & ... \\
$187$ & $                      \mathrm{Fe}^+(c^2G)+\mathrm{H}^-$ &   18& $   ^{2}\Sigma^+,\        ^{2}\Pi,\     ^{2}\Delta,\       ^{2}\Phi,\     ^{2}\Gamma$ \\
$188$ & $                      \mathrm{Fe}^+(b^2D)+\mathrm{H}^-$ &   10& $   ^{2}\Sigma^+,\        ^{2}\Pi,\     ^{2}\Delta$ \\
$189$ & $                      \mathrm{Fe}^+(a^2S)+\mathrm{H}^-$ &    2& $   ^{2}\Sigma^+$ \\
$190$ & $                      \mathrm{Fe}^+(c^2D)+\mathrm{H}^-$ &   10& $   ^{2}\Sigma^+,\        ^{2}\Pi,\     ^{2}\Delta$ \\
$191$ & $                     \mathrm{Fe}^+(d^2D2)+\mathrm{H}^-$ &   10& $   ^{2}\Sigma^+,\        ^{2}\Pi,\     ^{2}\Delta$ \\
&&&\\ 
     \multicolumn{3}{l}{Number of symmetries to calculate :  18} & $    ^{6}\Sigma^+,\        ^{6}\Pi,\     ^{6}\Delta,\   ^{4}\Sigma^-,\        ^{4}\Pi,\     ^{4}\Delta,\       ^{4}\Phi,\   ^{4}\Sigma^+,\   ^{2}\Sigma^+,\        ^{2}\Pi,\     ^{2}\Delta,\       ^{2}\Phi,\     ^{2}\Gamma,\   ^{2}\Sigma^-,\       ^{2}\mathrm{H},\     ^{4}\Gamma,\       ^{4}\mathrm{I},\      ^{2}\mathrm{I}$ \\
 \multicolumn{3}{l}{$g_\mathrm{total}:$} & $              6,\             12,\             12,\              4,\              8,\              8,\              8,\              4,\              2,\              4,\              4,\              4,\              4,\              2,\              4,\              8,\              8,\              4$ \\

\hline 
\end{tabular}
\end{table*}

\section{Results and discussion}

In simpler atoms, the core corresponding to the ground state of the ion is usually completely dominant because it leads to curve crossings at a range of internuclear distances, including those optimal for large transition probabilities at intermediate internuclear distances ($\sim$~30 to 50~au).  This is not the case for complex atoms such as iron, since there are low-lying states of \ion{Fe}{ii}, and thus other low-lying cores also make contributions.  However, the importance certainly decreases as the excitation of the core increases, thus leading more and more only to crossings with low-lying covalent states at shorter and shorter range.   For this reason we take examples from the two lowest cores, $\mathrm{Fe}^+(a^6D)$ and $\mathrm{Fe}^+(a^4F)$, which lead to the two lowest lying ionic states.  The $\mathrm{Fe}^+(a^6D)$ core leads to three symmetries, $^{6}\Sigma^+$, $^{6}\Pi$, and $^{6}\Delta$, and the $\mathrm{Fe}^+(a^4F)$ core leads to four symmetries, $^{4}\Sigma^-$, $^{4}\Pi$, $^{4}\Delta$, and $^{4}\Phi$.    Both cores have significant statistical weight,  thus strengthening their importance.   

\begin{figure}
\centering
\begin{overpic}[width=0.49\textwidth]{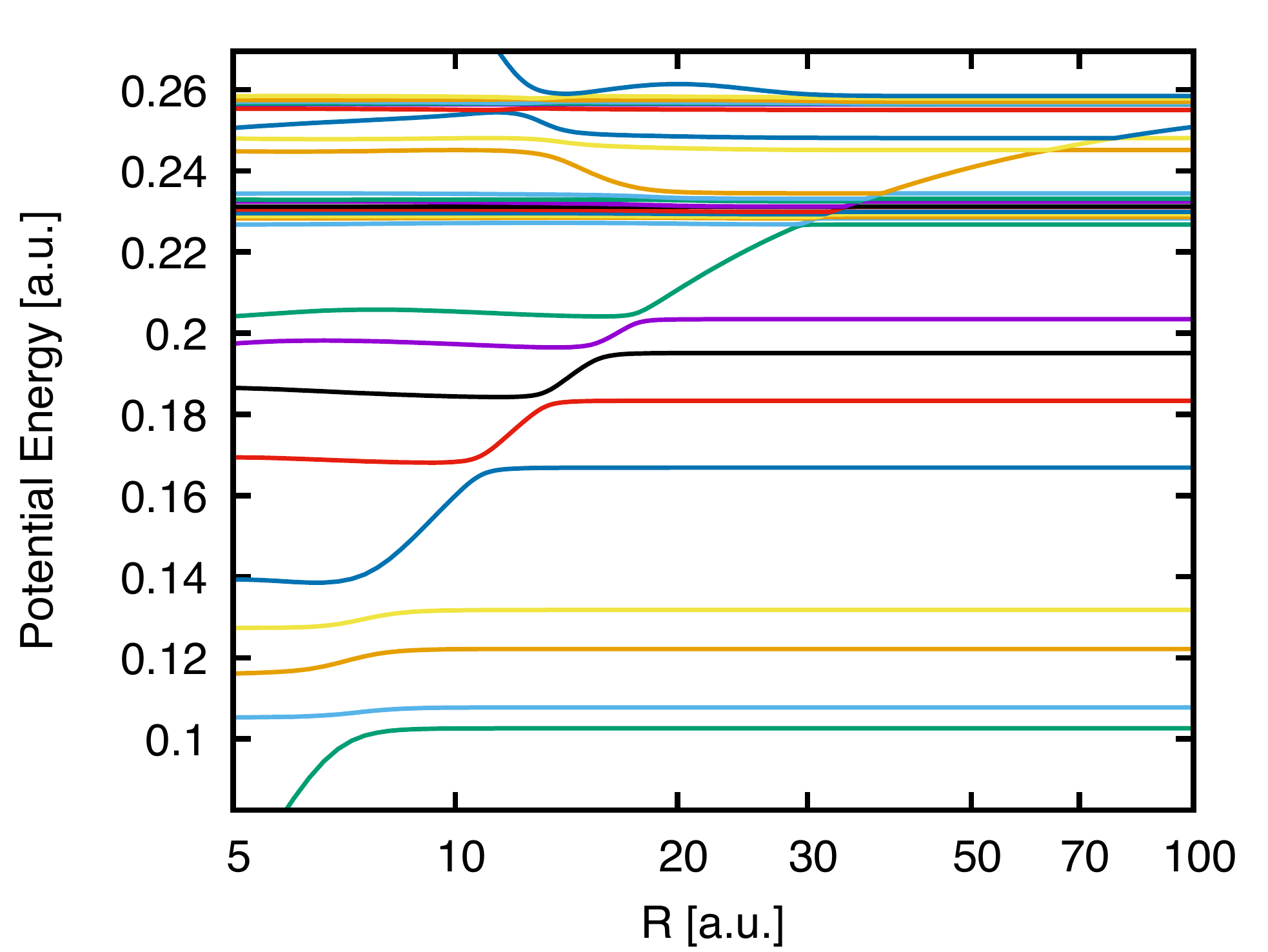}\put(40,30){$^6\Sigma^+$, $\mathrm{Fe}^+(a^6D)$ core}\end{overpic}
\begin{overpic}[width=0.49\textwidth]{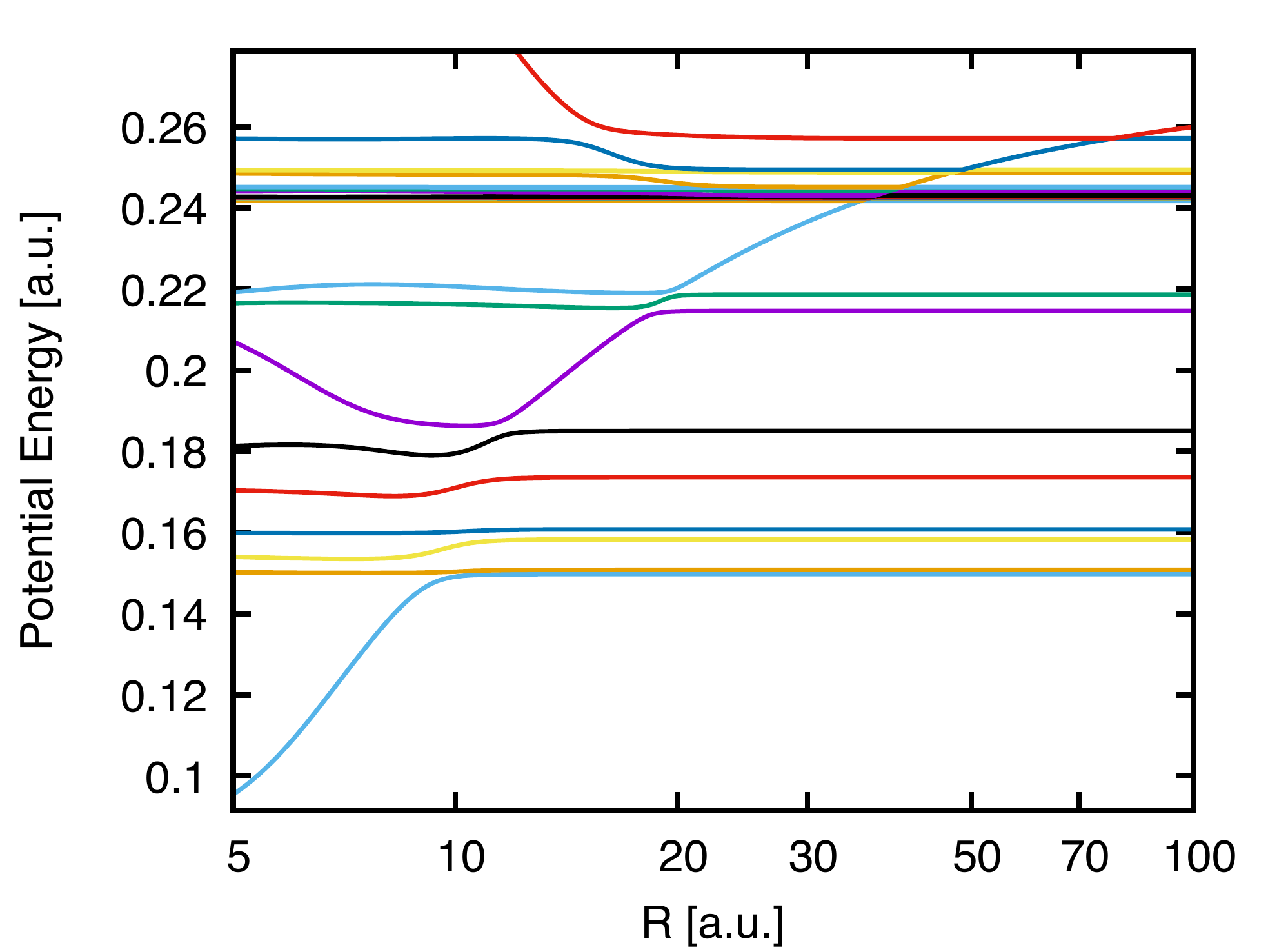}\put(40,20){$^4\Sigma^-$, $\mathrm{Fe}^+(a^4F)$ core}\end{overpic}
\caption{Example potentials energies for Fe+H from the LCAO model.  Upper panel: $^6\Sigma^+$ with $\mathrm{Fe}^+(a^6D)$ core. Lower panel: $^4\Sigma^-$ with $\mathrm{Fe}^+(a^4F)$ core. }\label{fig:pots}
\end{figure}

\begin{figure}
\centering
\begin{overpic}[width=0.49\textwidth]{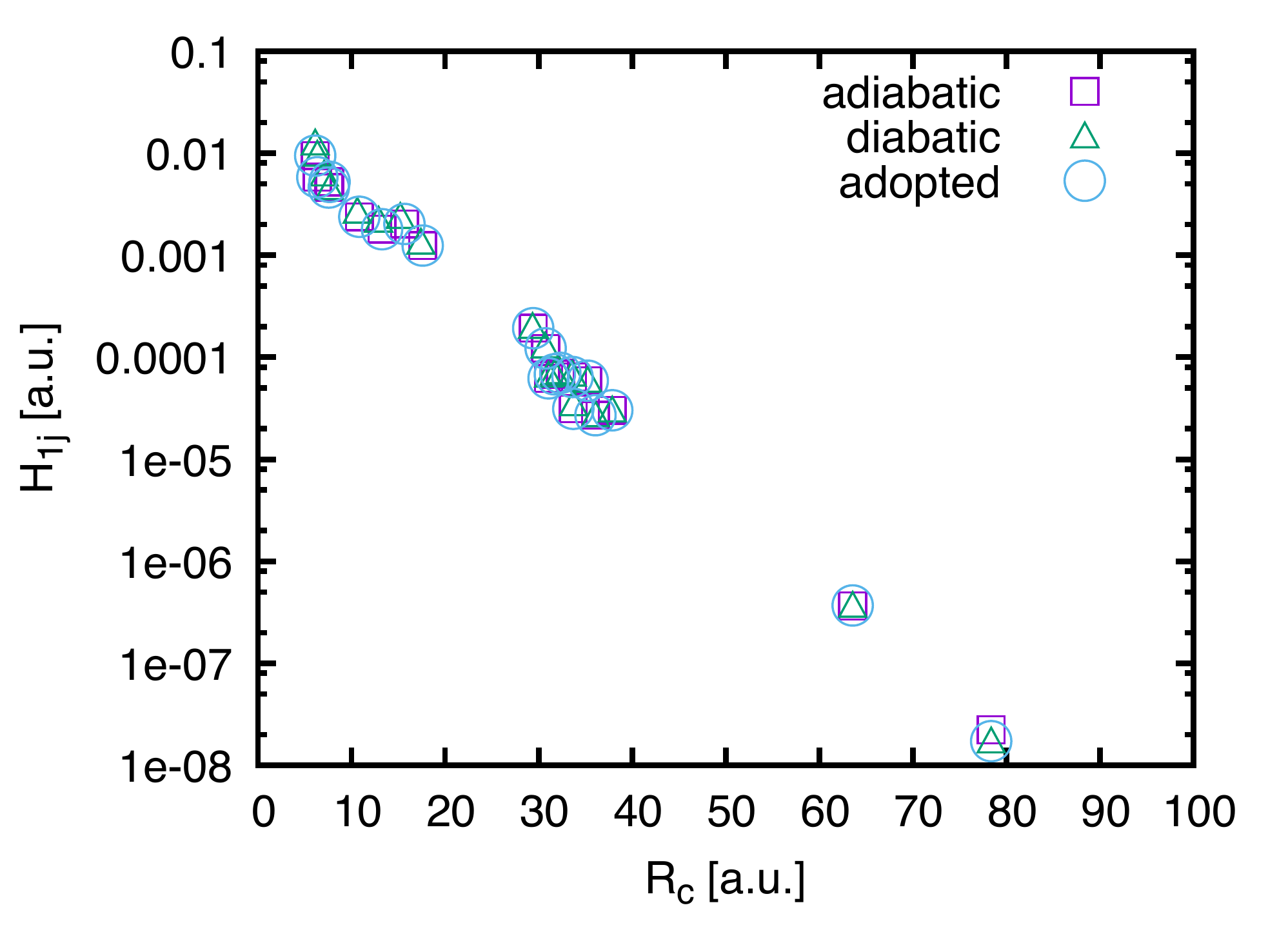}\put(30,25){$^6\Sigma^+$, $\mathrm{Fe}^+(a^6D)$ core}\end{overpic}
\begin{overpic}[width=0.49\textwidth]{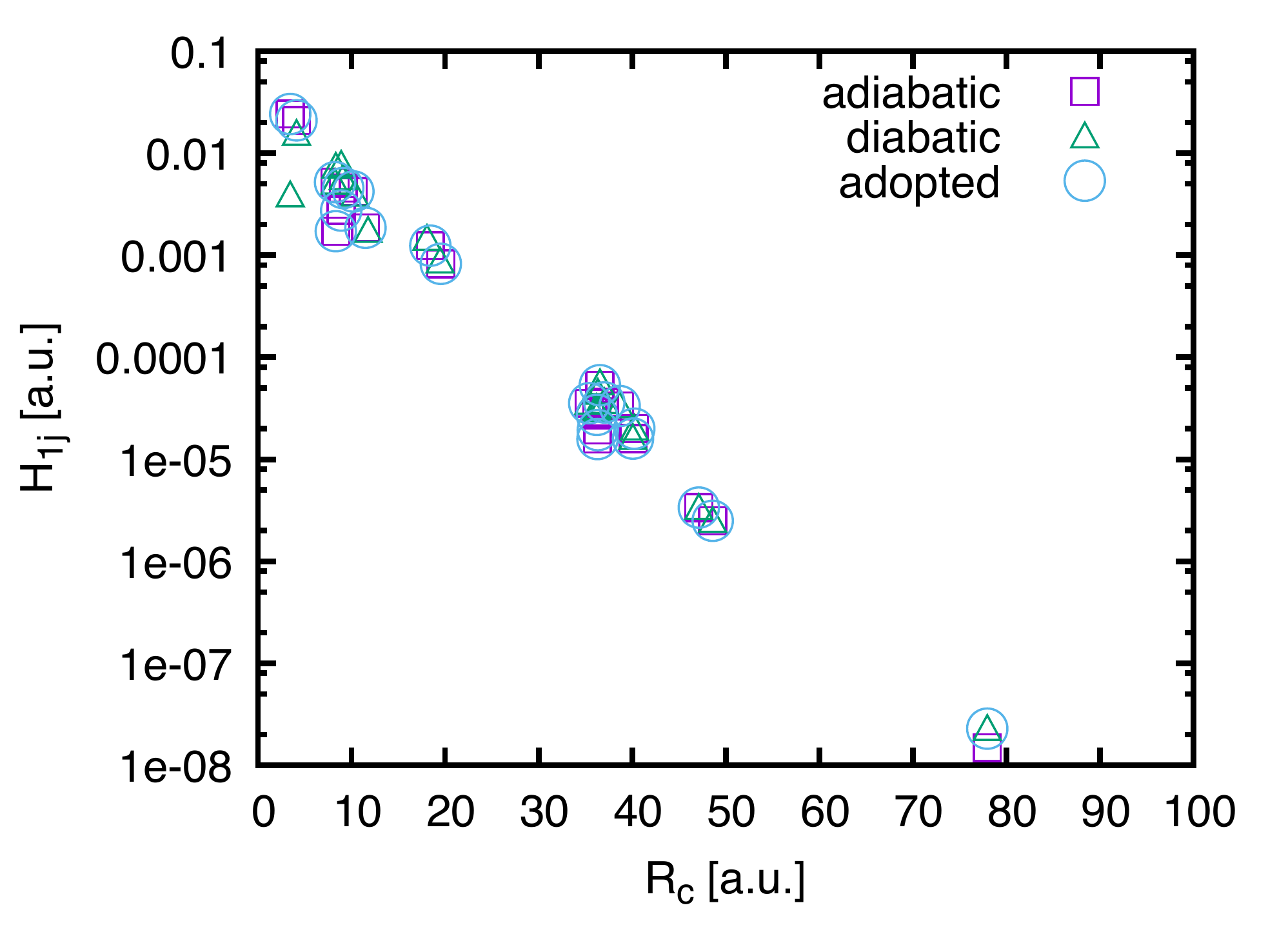}\put(30,25){$^4\Sigma^-$, $\mathrm{Fe}^+(a^4F)$ core}\end{overpic}
\caption{Example couplings $H_{1j}$ from the LCAO model plotted against the crossing distance $R_c$ for Fe+H states.  Upper panel: $^6\Sigma^+$ with $\mathrm{Fe}^+(a^6D)$ core. Lower panel: $^4\Sigma^-$ with $\mathrm{Fe}^+(a^4F)$ core. Results are shown for the adiabatic and diabatic models and the final adopted values. }\label{fig:lz}
\end{figure}

Figure~\ref{fig:pots} shows the $^{6}\Sigma^+$ potentials resulting from the $\mathrm{Fe}^+(a^6D)$ core and the $^{4}\Sigma^-$ potentials resulting from the $\mathrm{Fe}^+(a^4F)$ core, where the series of ionic curve crossings are clearly seen.  Figure~\ref{fig:lz} shows the corresponding derived ionic-covalent couplings $H_{1j}$ at the avoided crossings, derived in the adiabatic and diabatic representations (see {\S}B of B16), which are in good agreement, except for a few cases at very short range.  These cases are unimportant as they cannot lead to large cross sections, and the adopted value from the adiabatic representation is used as it is expected to be more reliable; see B16.  For the $^{6}\Sigma^+$ $\mathrm{Fe}^+(a^6D)$ core case it is seen that there are essentially two clusters of crossings: the lowest lying covalent states with crossings at short distance ($\sim 10$ to 20~au) correspond to Fe states with excitations of 0.1 to 0.2~au ($\sim 24000$~cm$^{-1}$ to $40000$~cm$^{-1}$ $\sim$ 3 to 5~eV) arising from configuration $3d^6(^5D)4s(^6D)4p$, and another large, tightly bunched group of somewhat excited states with crossings at intermediate internuclear distances ($\sim 30$ to 40~au) correspond to Fe states with excitations of 0.22 to 0.24 ($\sim 50000$~cm$^{-1}$ $\sim$ 6.3~eV) arising from configurations $3d^6(^5D)4s(^6D)4d$ and $3d^6(^5D)4s(^6D)5p$.  Similarly, for the $^{4}\Sigma^-$ $\mathrm{Fe}^+(a^4F)$ core case, two clusters are also seen, but do not correspond to the same groups of states.  The lowest lying covalent states with crossings at short distance ($\sim 10$ to 20~au) correspond to Fe states with excitations of 0.14 to 0.22~au ($\sim 31000$~cm$^{-1}$ to $48000$~cm$^{-1}$ $\sim$ 3.8 to 6~eV) arising predominantly from $3d^7(^4F)4p$, and an additional tightly bunched group of many excited states with crossings at intermediate internuclear distances ($\sim 30$ to 40~a.u.) correspond to Fe states with excitations of 0.24 to 0.25 ($\sim 53000$~cm$^{-1}$ $\sim$ 6.6~eV) arising from $3d^7(^4F)4d$ and $3d^7(^4F)5p$.  

\begin{figure*}
\centering
\begin{overpic}[abs,unit=1mm,grid=false,tics=5]{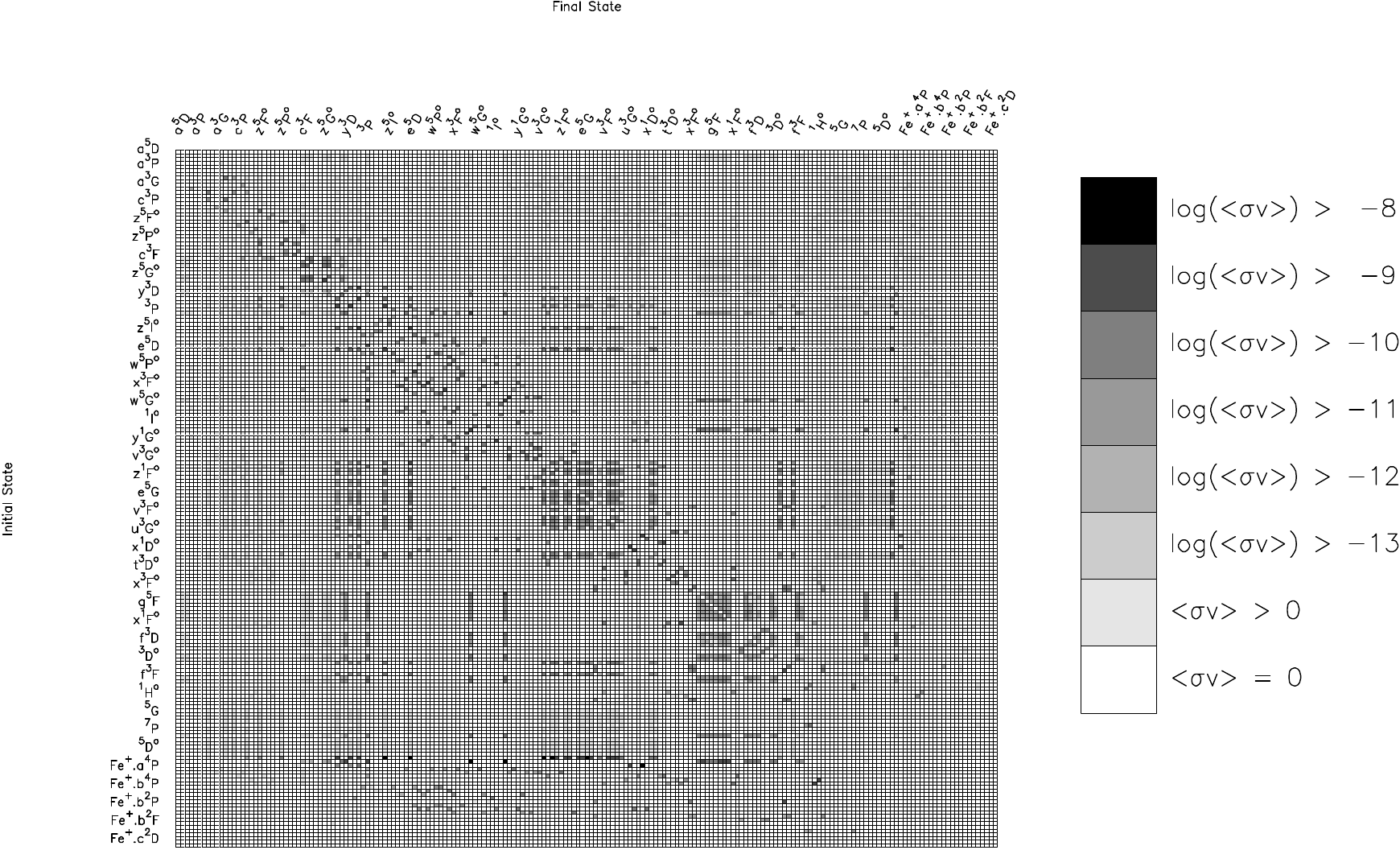}
\linethickness{0.2mm} \color{blue}%
\put(0,0){\polygon(21.2,0.2)(21.2,11.9)(111.4,11.9)(111.4,0.2)}
\color{red}
\put(0,0){\polygon(111.4,11.9)(111.4,88.2)(125.0,88.2)(125.0,11.9)}
\end{overpic}
\caption{Graphical representation of the rate coefficient matrix  $\langle \sigma \varv \rangle$ (in cm$^3$~s$^{-1}$) for inelastic Fe + H and Fe$^+$ + H$^-$ collisions at temperature $T = 6000$~K.  Results are from the LCAO asymptotic model.  The logarithms in the legend are to base 10.  Only every fifth state is labelled.  The charge transfer processes, involving initial or final ionic states, are outlined with (coloured) boxes, for ion-pair production at the upper right (red) and for mutual neutralisation at the lower left (blue).}\label{fig:rates_grid}
\end{figure*}

\begin{figure}
\centering
\includegraphics[width=0.49\textwidth]{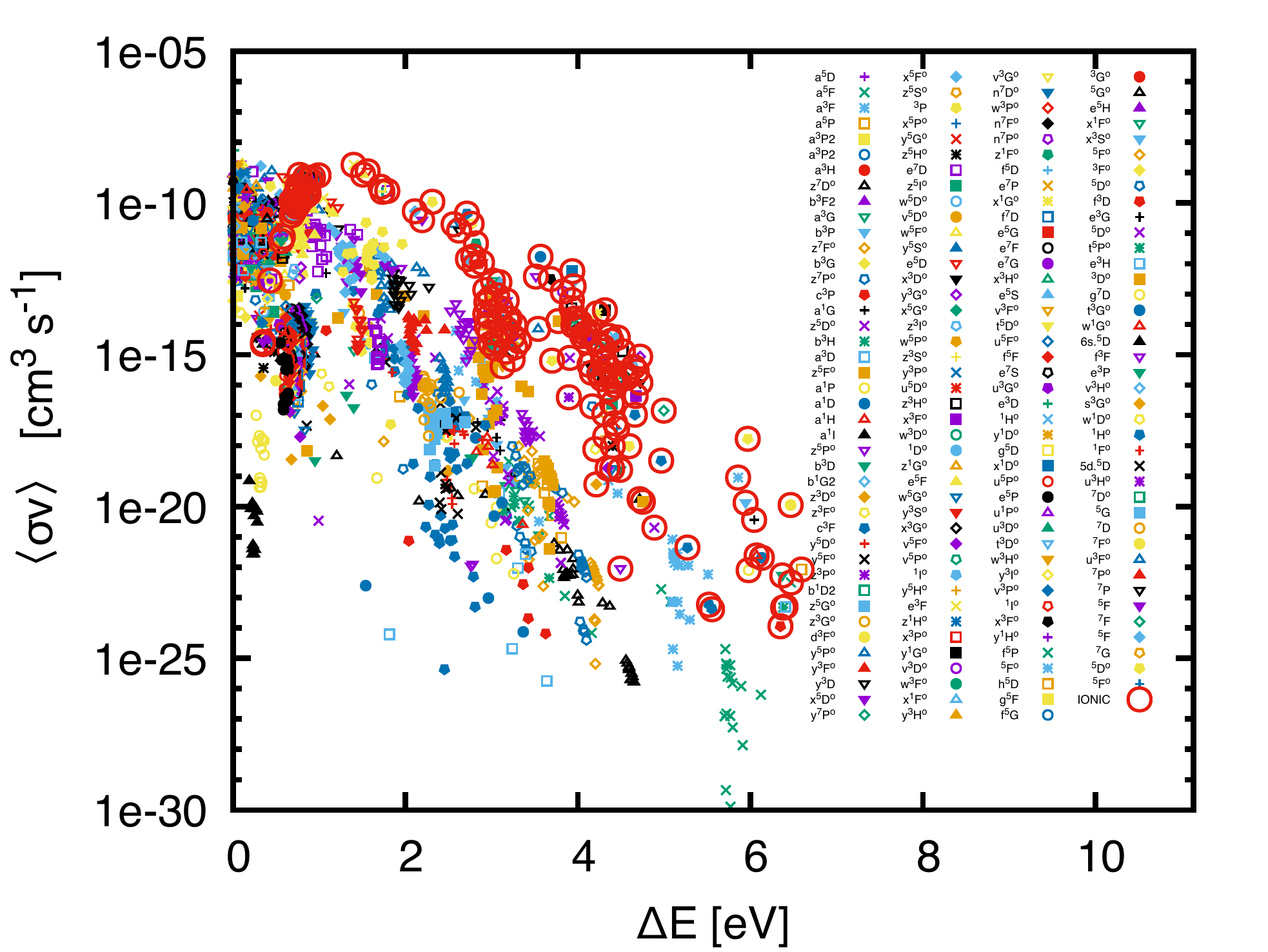}
\caption{Rate coefficients $\langle \sigma \varv \rangle,$ for Fe+H collision processes at 6000~K, plotted against the asymptotic energy difference between initial and final molecular states, $\Delta E$.  The data are shown for endothermic processes. i.e. excitation and ion-pair production.  The legend labels the initial state of the transition and processes leading to a final ionic state (ion-pair production) are circled. }\label{fig:rates}
\end{figure}

Example final results for the rate coefficients are shown in two forms in Figs.~\ref{fig:rates_grid} and \ref{fig:rates}.  Figure~\ref{fig:rates_grid} is particularly useful for identifying the dominant cores, as these have large rates for mutual neutralisation.  As expected, and discussed above, the core corresponding to the ground state of $\mathrm{Fe}^+(a^6D)$ gives rise to the most efficient charge transfer (mutual neutralisation and ion pair production) processes of the form  $\mathrm{Fe}^+(a^6D) + \mathrm{H}^- \rightleftarrows \mathrm{Fe} + \mathrm{H}$, as these show the largest rate coefficients.  
The processes are to and from the Fe states corresponding to the cluster of states at around $\sim 0.23$~au, $\sim 50000$~cm$^{-1}$ $\sim 6.3$~eV.  The first excited core, $\mathrm{Fe}^+(a^4F)$, also provides some significant rate coefficients, but these coefficients are smaller than those for the ground core and are shifted towards the more excited states (around 0.24~au, 53000~cm$^{-1}$, 6.6~eV).   The remaining cores show some contributions to charge transfer processes, but are typically orders of magnitude lower.  

Fig.~\ref{fig:rates} shows the endothermic rates as a function of change in energy state for the system, $\Delta E$.  Earlier calculations for simple atoms, have all shown a rather regular behaviour of the rate coefficients for charge transfer processes with $\Delta E$, that is the asymptotic energy difference between initial and final molecular states.  The rate coefficients for ion-pair production usually show an arch, peaking at transitions around $\Delta E=1$~eV \citep[e.g.][]{Barklem2012}, corresponding to transitions resulting from crossings at optimal internuclear distance, i.e. $\sim$~30 to 50~au, corresponding to crossings about 1~eV ($\sim 0.04$~au) below the asymptotic limit for the ionic state.  \cite{ezzeddine_role_2016} have exploited this regular behaviour by fitting the data for simple atoms to permit extrapolation to complex atoms such as Fe.  The same structure for the ion-pair production processes is seen in Fig.~\ref{fig:rates}, although additional structure is seen at $\Delta E=3$--5~eV, where rather than a monotonic decrease with  $\Delta E$, two branches are seen.  This is due to the complicated interplay of the influence of more than one core and the clustering of states correlating to different core configurations.  This structure only appears for processes with rather small rates.

As in previous work (B16, \citealt{BarklemExcitationchargetransfer2017}) calculations were also performed with alternate models for the couplings, for example the SEMI-EMP and LH-J models of B16, which provide fluctuations and an indication of the uncertainties, at least for the largest rates for which the ionic crossing mechanism is dominant.  These fluctuations are not shown in Fig.~\ref{fig:rates}, as the plot becomes too dense.  However, the fluctuations of the rate coefficients show very similar behaviour to that for other atoms studied in earlier work, namely, roughly one order of magnitude fluctuations for the largest rates and larger fluctuations for smaller rates.  

\section{Concluding remarks}

Large-scale model calculations have been performed for inelastic Fe+H collision processes using a model approach accounting for the ionic crossing mechanism associated with electron transfer between ionic and covalent configurations.  The calculations are expected to give reasonable estimates for the processes with the largest rates, particularly charge transfer and excitation processes between near-lying states.  The results show general behaviour very similar to that seen for earlier calculations with simple atoms, although some additional complexity is seen owing to the influence of more than a single core configuration for the iron atom.  In general it is found that the largest rates are for charge transfer processes to and from the two clusters of states around 6.3 and 6.6~eV, corresponding in both cases to active $4d$ and $5p$ electrons undergoing transfer.  The excitation and de-excitation transition rates are largest for transitions between nearby states; these clusters are extensively coupled.

Uncertainties for the largest rates are expected to be about one order of magnitude.  The uncertainties for processes with smaller rates are larger, both because of the uncertainties in the present model and uncertainties from mechanisms not captured by the present model.  It should be noted that recent work on comparing the centre-to-limb variation of the solar 777~nm oxygen lines, perhaps indicates the need for larger efficiency of hydrogen collision processes for excitation among low-lying states, and it seems possible that this could be provided by mechanisms other than the ionic crossing mechanism (Amarsi et~al.\ in preparation).  This may well be the case for other complex atoms such as iron, and will be the subject of future research, both in terms of comparison with centre-to-limb variation in the sun and theoretical calculations.  Previous work on the solar centre-to-limb variation for Fe by \cite{lind_non-lte_2017} made use of preliminary calculations using this method.  The present calculation corrects an earlier error \citep{barklem_erratum:_2017} and includes a number of other improvements \citep{BarklemExcitationchargetransfer2017} and many more states and cores.

\begin{acknowledgements}
This work received financial support from the Swedish Research Council and the project grant ``The New Milky Way'' from the Knut and Alice Wallenberg Foundation.
\end{acknowledgements}

\bibliographystyle{aa} 
\bibliography{MyLibrary.bib}

\end{document}